# Title: Direct observation of van der Waals stacking dependent interlayer magnetism

**Authors:** Weijong Chen[1], Zeyuan Sun[1], Lehua Gu[1], Xiaodong Xu[2], Shiwei Wu[1,3*], Chunlei Gao[1,3*]

**Affiliations:**

[1] State Key Laboratory of Surface Physics, Key Laboratory of Micro and Nano Photonic Structures (MOE), and Department of Physics, Fudan University, Shanghai 200433, China.

[2] Department of Physics, and Department of Materials Science and Engineering, University of Washington, Seattle, Washington 98195, USA.

[3] Collaborative Innovation Center of Advanced Microstructures, Nanjing 210093, China.

*Corresponding emails: clgao@fudan.edu.cn, swwu@fudan.edu.cn.

**Abstract: Controlling the crystal structure is a powerful approach for manipulating the fundamental properties of solids. Unique to two-dimensional (2D) van der Waals materials, the control can be achieved by modifying the stacking order through rotation and translation between the layers. Here, we report the first observation of stacking dependent interlayer magnetism in the 2D magnetic semiconductor, chromium tribromide ($CrBr_3$), enabled by the successful growth of its monolayer and bilayer through molecular beam epitaxy. Using in situ spin-polarized scanning tunneling microscopy and spectroscopy, we directly correlated the atomic lattice structure with observed magnetic order. We demonstrated that while individual $CrBr_3$ monolayer is ferromagnetic, the interlayer coupling in bilayer depends strongly on the stacking order and can be either ferromagnetic or antiferromagnetic. Our observations provide direct experimental evidence for exploring the stacking dependent layered magnetism, and pave the way for manipulating 2D magnetism with unique layer twist angle control.**

**Main Text:**

Van der Waals (vdW) stacking has been extensively recognized as a critical component in determining the properties of layered vdW materials. In particular, weak interlayer vdW interactions allow for the control over the rotational and translational degrees of freedom between layers, creating a host of new materials with distinct stacking symmetries and functionalities([1-3]). While most previous work has focused on the electronic and optical properties associated with the vdW stacking([4-7]), the recent discovery of magnetism in 2D materials, achieved through both mechanical exfoliation([8-13]) and molecular beam epitaxy (MBE) ([14], [15]), provides an exciting opportunity to explore the effects of stacking order on a material's magnetic properties.

Among the 2D magnetic materials discovered so far, the family of chromium trihalides, $CrX_3$ (X = Cl, Br, I), has garnered special interest([8], [16]). For instance, ferromagnetism perpendicular to the 2D plane persists in $CrI_3$ monolayers, while its bilayers exhibit interlayer antiferromagnetism. These magnetic structures lead to a number of emerging phenomena, such as giant tunneling magnetoresistance in spin-filter magnetic tunnel junctions([17-19]) and the electrical control of 2D magnetism([20-22]). In contrast to $CrI_3$, recent tunneling measurements suggest that the interlayer coupling in atomically thin $CrBr_3$ is ferromagnetic([23]). This is confirmed through polar reflectance magneto circular dichroism (RMCD) measurements of mechanically exfoliated bilayers of $CrBr_3$, which reveal magnetic hysteresis behavior centered around zero applied magnetic field, distinct from the three-step staircase pattern observed in bilayer $CrI_3$ (Supplementary Materials Fig. S1). Given that $CrI_3$ and $CrBr_3$ are isostructural, such a significant difference in the magnetic properties between bilayers of $CrI_3$ and $CrBr_3$ prompts a thorough investigation of the mechanisms governing the interlayer magnetic coupling.

In this work, using in situ spin-polarized scanning tunneling microscopy and spectroscopy ([24], [25]), we establish a direct correlation between the interlayer magnetic coupling and the stacking structures in $CrBr_3$ bilayer. The $CrBr_3$ films were grown on freshly cleaved highly oriented pyrolytic graphite (HOPG) substrates by molecular beam epitaxy (MBE). During the growth, the sample surface was monitored *in situ* by reflection high-energy electron diffraction (RHEED). Figures 1A and B show the RHEED patterns before and after growing $CrBr_3$ with marked diffraction orders, respectively. The stripe-like RHEED pattern from the $CrBr_3$ indicates the formation of a 2D crystalline thin film, confirmed to be a $CrBr_3$ monolayer by scanning tunneling microscopy (STM) (Fig. 1C). Further deposition of $CrBr_3$ led to the appearance of $CrBr_3$ bilayer islands (Fig. 1D) (for details about the MBE growth, see Materials and Methods).

Figure 1E shows the atomically resolved STM image of the $CrBr_3$ monolayer, revealing periodically spaced triangular cluster protrusions. To understand this image, we consider the monolayer crystal structure of $CrBr_3$. As shown in Fig. 1F, Cr atoms are arranged in a honeycomb lattice structure and each atom is surrounded by an octahedron of six Br atoms([16]). Within a single honeycomb formed by six Cr atoms, there are three Br atoms at the top and bottom surfaces, marked by solid and dotted green triangles with opposite orientations, respectively. Thus, the $CrBr_3$ monolayer has three-fold rotational symmetry, with the rotational axis across either the Cr atoms or the center of the honeycomb. By overlaying the atomic structure on top of the STM image in Fig. 1E, we determined that the observed triangular cluster is formed by the three top Br atoms (solid green triangle in Fig. 1E). The measured in-plane lattice constant is 6.3 Å, consistent with the bulk lattice constant (6.26 Å) ([26]). The monolayer thickness is determined to be 6.5 Å by atomic force microscopy (AFM), displayed in Fig. 1G.

Structural domains and their boundaries are also clearly seen in the image. Both the large-scale topography and atomically resolved STM images demonstrate the high-quality growth of the CrBr$_3$ monolayer films.

To measure the magnetic properties, we carried out *in situ* spin-polarized STM measurements with an applied out-of-plane magnetic field at 5 K. An antiferromagnetic Cr-coated tungsten tip was used, whose magnetization at the tip apex does not change under magnetic fields below 2 T(*27*). Figure 2A shows the dI/dV spectra from a CrBr$_3$ monolayer with oppositely applied fields (B = ±0.3 T), measured at the same position. While the two dI/dV curves show the semiconducting behavior near zero bias voltage, they clearly differ with bias voltage $V_b$ over 1 V, which results from the magnetization flip of the CrBr$_3$ monolayer under the reversal of the external magnetic field. To further confirm the existence of ferromagnetism, we measured the dI/dV signal at $V_b$ = 1.4 V while sweeping the magnetic field back and forth, as shown in Fig. 2B. A rectangular hysteresis loop was observed with a coercive field of ~30 mT, a hallmark of ferromagnetism in the CrBr$_3$ monolayer. Thus, we conclude that epitaxial CrBr$_3$ monolayers grown on HOPG are semiconducting ferromagnets with a well-defined out-of-plane easy axis.

With the atomic structure of CrBr$_3$ monolayer determined and its ferromagnetism confirmed, we now focus on CrBr$_3$ bilayers. In general, there are two types of stacking structures with the unit cell size the same as that of the monolayer: R-type with both layers aligned to the same orientation, and H-type with a 180° rotation between the layers. For R-type stacking, two polytypes for CrBr$_3$ bulk crystals were reported: rhombohedral at low temperature and monoclinic at high temperature (above 420K)(*16*). These two structures mainly differ on the relative translation between the two monolayers. However, common to vdW materials(*4, 5, 28*), the energy variation between different stacking structures can be relatively small, even if the stacking structure is absent in bulk crystal. Therefore, the possible stacking structures in the bilayer case could be far more complex than those reported in literature. By only considering the high symmetry configurations (i.e., hollow, bridge and top sites) between the halogen atoms at the interface of adjacent layers, there are already six and five polytypes for R-type stacking (summarized in Supplementary Materials Table S1) and H-type stacking (Supplementary Materials Table S2), respectively.

We found that the H- and R-type stacking structures can both form in the MBE grown bilayers, and they give rise to distinct interlayer magnetic coupling. The results corresponding to the H-type stacking structure are given in Fig. 3. Figure 3A shows the STM image across the step edge between a CrBr$_3$ bilayer island and a CrBr$_3$ monolayer film which is an extension of the bilayer's bottom layer. Their atomically resolved STM images are shown in Figs. 3B and C, which show individual triangular clusters formed by the top three Br atoms in the unit cell of CrBr$_3$, similar to those in Fig. 1E. The orientation of the triangular clusters from the monolayer region, which has the same crystal structure as the bottom layer of the bilayer, is exactly opposite to the top layer in bilayer, suggesting H-type stacking.

To determine the translational degree of freedom between the two layers, we overlaid the two CrBr$_3$ lattices, represented by the magenta and green triangles, on the top and bottom layers of the bilayer island in Fig. 3A (further details given in Extended Data Fig. 2). We found that the top layer is translated to 0.55**a**+0.20**b**, with respect to the bottom layer (Fig. 3D). Given that the experimental uncertainty of translation is about ±7% of the lattice constants ***a*** and ***b***, the closest stacking structure is "Bridge I" of H-type stacking listed in Supplementary Materials Table S2.

In terms of both rotational and translational degrees of freedom, this observed stacking structure is distinct from the reported structure models for $CrX_3$ in the literature([16, 26]). Further spin-polarized STM measurement on this H-type stacked $CrBr_3$ bilayer showed a rectangular magnetic hysteresis loop, with coercive magnetic field of ~45 mT (Fig. 3E), similar to the hysteresis observed in $CrBr_3$ monolayers (Fig. 2B). Thus, in H-type stacked $CrBr_3$ bilayers, the interlayer coupling is ferromagnetic.

The second kind of stacking structure is depicted in Fig. 4, where the triangular clusters from the adjacent monolayer and bilayer regions are oriented in the same direction (Fig. 4A-C), suggesting R-type stacking. The translational degree of freedom between the two layers is determined in the same way as in Fig. 3. In R-type stacking, the top layer is translated to 0.48**a**+0.48**b**, with respect to the bottom layer (Fig. 4D). This stacking structure is also distinct from the reported rhombohedral and monoclinic structures([16]) or any other high symmetry configurations for $CrX_3$. Rather, it takes a position in the middle of the hollow and top sites for halogen atoms at the interface, which is named "Special" listed in Supplementary Materials Table S1.

For this R-type stacked bilayer, the magnetic field dependent dI/dV signal could be directly observed with a nonmagnetic tungsten tip. Starting with zero applied magnetic field, sweeping the external magnetic field in either direction results in a plateau in the dI/dV curve. Beyond magnetic fields of ±0.5 T, the dI/dV abruptly increases (Fig. 4E). Such behavior in the dI/dV signal is identical to the giant tunneling magnetoresistance recently discovered in spin-filter magnetic tunnel junctions using mechanically exfoliated $CrI_3$ bilayers([17-19]). This strongly suggests that the interlayer coupling of R-type stacked $CrBr_3$ bilayers is antiferromagnetic. The anti-aligned spin filters for |B| < 0.5 T suppress the dI/dV signal at $V_b$ = 0.825 V, while large magnetic fields beyond ±0.5 T align the magnetization between the layers and thus increases the dI/dV signal. This magnetic field-driven layered antiferromagnetic to ferromagnetic switching gives rise to the observed dI/dV contrast. In addition, this magnetic contrast is also reflected in the STM height (Supplementary Materials Fig. S4), which further confirms the antiferromagnetic ground state in this R-type stacked bilayer. For comparison, when a nonmagnetic tungsten tip was used for $CrBr_3$ H-type stacked bilayers that were ferromagnetically coupled, no magnetic field dependent dI/dV features could be observed (Supplementary Materials Fig. S5).

The distinct interlayer magnetism, from antiferromagnetic coupling in R-type stacking to ferromagnetic coupling in H-type stacking, in our MBE-grown $CrBr_3$ bilayers clearly demonstrates the remarkable tunability of 2D magnetism through the stacking order. The interlayer coupling in $CrBr_3$ bilayers is mediated through superexchange interaction, which is controlled by the directional hybridization between the Br p orbitals and Cr d orbitals([29, 30]). Since the bond angles as well as the bond distances of the Cr-Br-Br-Cr exchange path are strongly dependent on the stacking order, the interlayer magnetism is expected to also depend on the specific stacking structure, including the interlayer distance and atomic site positions. In fact, two recent theoretical papers have investigated the interlayer magnetism of the rhombohedral and monoclinic structures in $CrI_3$ bilayer([31, 32]).

In summary, we observed new stacking structures and their associated interlayer magnetic coupling in MBE-grown $CrBr_3$ bilayers. While the exact growth mechanism remains to be investigated, this observation illustrates the polytypism in vdW materials and its importance to 2D magnetism. Our work also calls for a close examination for the stacking structures in mechanically exfoliated $CrX_3$ samples, as the exfoliated $CrI_3$ and $CrBr_3$ bilayer exhibit distinct interlayer magnetic coupling. Nevertheless, it is a first step toward the microscopic understanding of interlayer magnetism governed by stacking order. We envision that this working principle could be used to manipulate 2D magnetism, for example, engineering spatially dependent spin textures uniquely enabled by twisted bilayers and heterostructures.

## ACKNOWLEDGMENTS

We thank Di Xiao and Ting Cao for insightful discussion, Bevin Huang for proofreading the paper and Donglai Feng for lending the AFM. **Funding:** The work at Fudan was supported by the National Natural Science Foundation of China (Grant Nos. 11427902, 11674063), National Key Research and Development Program of China (Grant Nos. 2016YFA0300904, 2016YFA0301002), and National Basic Research Program of China (Grant No. 2014CB921601). X.X. is supported by the Department of Energy, Basic Energy Sciences, Materials Sciences and Engineering Division (DE-SC0018171). **Author contributions:** C.G. and S.W. conceived and supervised the project. W.C. grew the sample and conducted the STM experiment. Z.S. conducted the RMCD measurement. L.G. and W.C. conducted the AFM measurement. W.C., Z.S., X.X., S.W. and C.G. analyzed the data and wrote the paper. **Competing interests:** The authors declare no competing interests. **Data and materials availability:** The data shown in the paper is available.

## Supplementary Materials:

Materials and Methods

Tables S1-S2

Figures S1-S5

**Figures and captions**

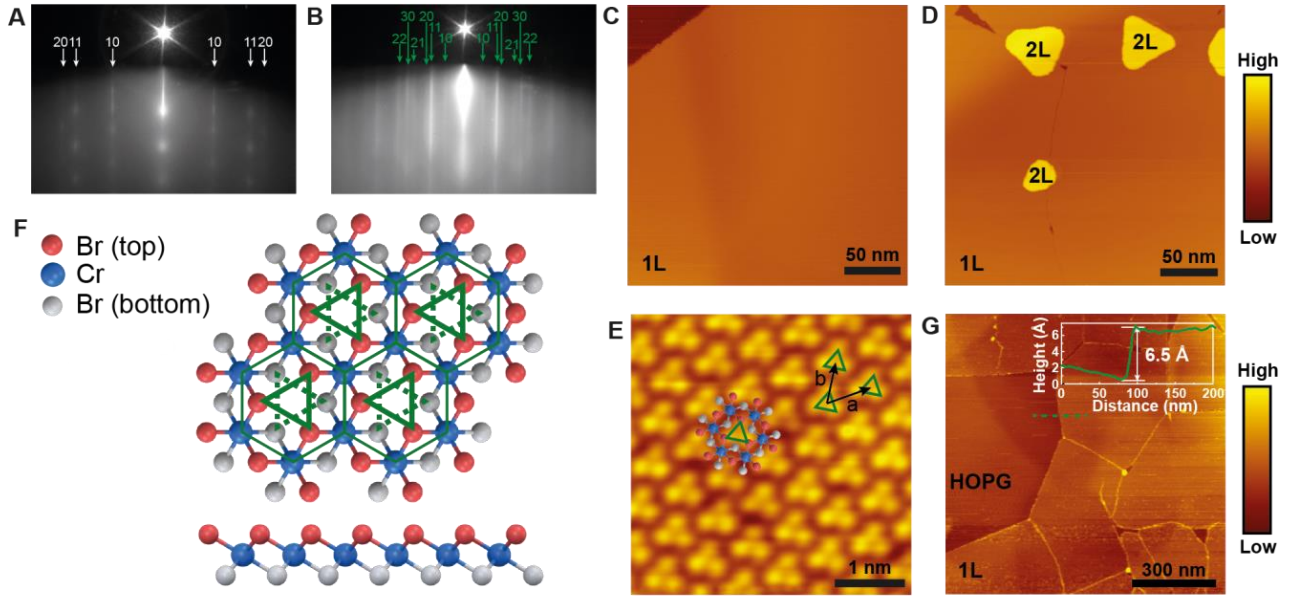

**Fig. 1. MBE Growth of CrBr₃ monolayer and bilayer on HOPG. A, B,** RHEED patterns with indicated diffraction orders of the bare HOPG substrate and the MBE-grown CrBr₃ film, respectively. **C, D,** STM images of CrBr₃ monolayer (**C**) with bilayer islands (**D**). The scan parameters were: $V_b$ = 1.3 V, $I$ = 100 pA, $T$ = 5 K. **E,** Atomically resolved image of a CrBr₃ monolayer with an overlaid atomic structure. $V_b$ = 1.7 V, $I$ = 500 pA, $T$ = 5 K. The lattice constants are determined to be 6.3 Å for the primitive vectors **a** and **b**, consistent with the bulk values. **F,** Top and side views of the monolayer CrBr₃ atomic structure. The Cr atoms form a honeycomb lattice sandwiched by Br atoms. Within the Cr honeycomb lattice, the top and bottom surfaces of Br atoms form single triangles but with opposite orientation, marked in solid and dotted green lines, respectively. **G,** AFM image of CrBr₃ monolayer with partial coverage. A line-cut profile across the monolayer and bare substrate is shown with a monolayer height of ~6.5 Å.

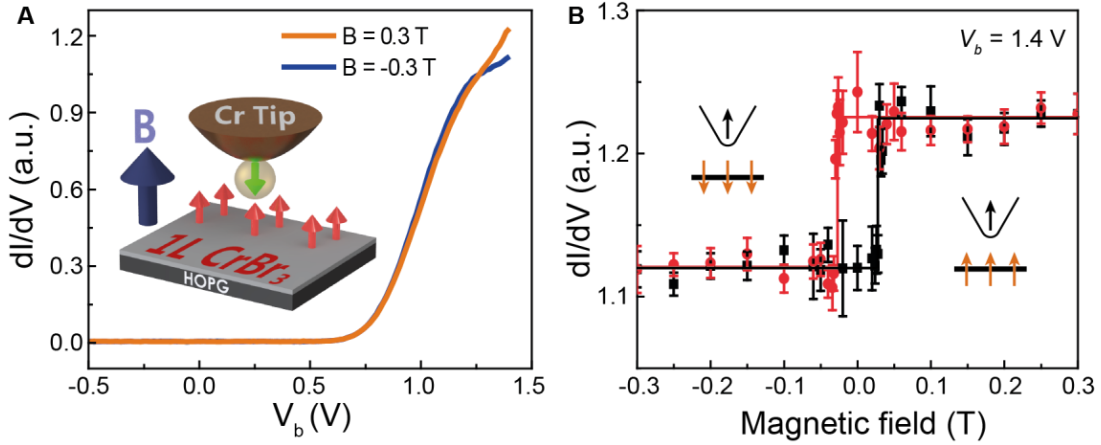

**Fig. 2. Spin-polarized tunneling of CrBr₃ monolayer. A,** Spin-polarized tunneling spectra under positive and negative out-of-plane magnetic fields (± 0.3 T). The inset illustrates the experimental geometry. **B,** The dI/dV signal as a function of the magnetic field. $V_b$ was fixed at 1.4 V. The out-of-plane magnetic field was swept upward (black data) and downward (red data). The ferromagnetic hysteresis loop is outlined as rectangular solid lines. Insets sketch the two configurations of the magnetization alignment between the Cr tip and the CrBr₃ monolayer film. Note that the in-plane component of magnetization in the Cr tip, if any, does not contribute to the magnetic contrast in dI/dV.

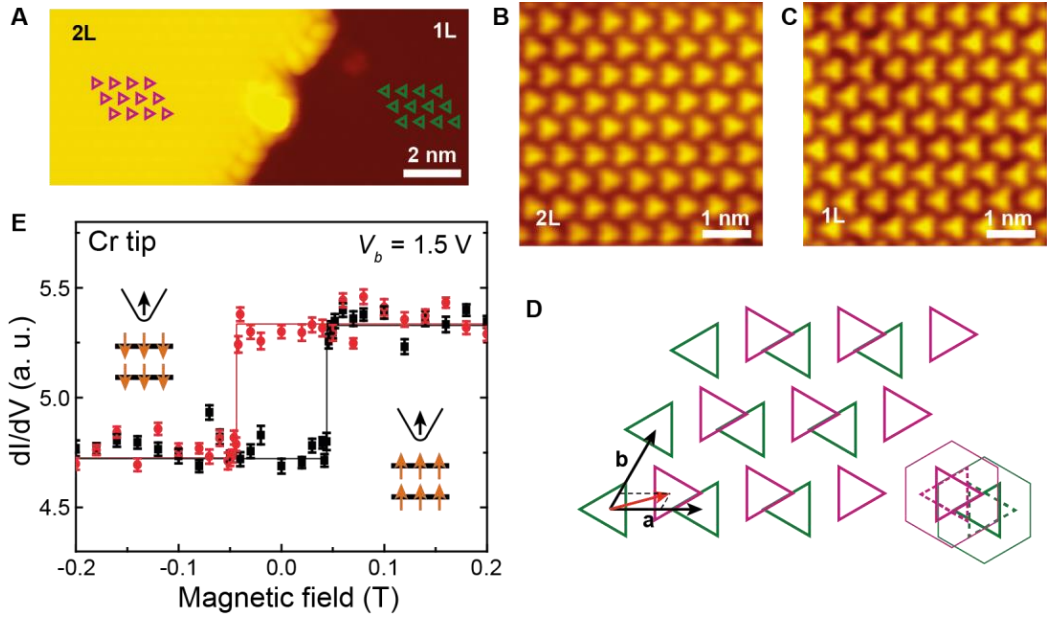

**Fig. 3. Interlayer ferromagnetic coupling in an H-type stacked CrBr₃ bilayer. A,** STM image of a CrBr₃ film with both a monolayer (1L) region and a bilayer (2L) island. **B, C,** Zoomed in atomically resolved images of the bilayer region (**B**) and its extended bottom monolayer (**C**) at $V_b$ = 1.9 V, identifying that the top and bottom layers in the bilayer are anti-aligned, or rotated by 180° (H-type stacking). **D,** Atomic structure of the CrBr₃ bilayer, as determined from atomically resolved STM scans (see Supplementary Materials Fig. S2 for further details). The unit cells of the top and bottom layers are represented by the magneta and green solid triangles, corresponding to the top surface of Br atoms in each monolayer sheet. These magneta and green solid triangles are also overlaid on the monolayer and bilayer in **A**. The unit cell of the top layer (magneta) is translated to 0.55**a** + 0.20**b** of the bottom layer (green). To compare with the structures in Supplementary Materials Table S2, the stacking structure is also shown with the bottom surface of Br atoms of each monolayer sheet in dotted triangles, and the Cr atoms in solid hexagons. **E,** Spin-polarized tunneling on the CrBr₃ bilayer as a function of magnetic field with a Cr-coated W tip at $V_b$ of 1.5 V. Similar to that of the CrBr₃ monolayer, a rectangular ferromagnetic hysteresis loop was observed with a coercive field of ~45 mT. Insets depict two configurations of the magnetization alignment between the tip and the sample.

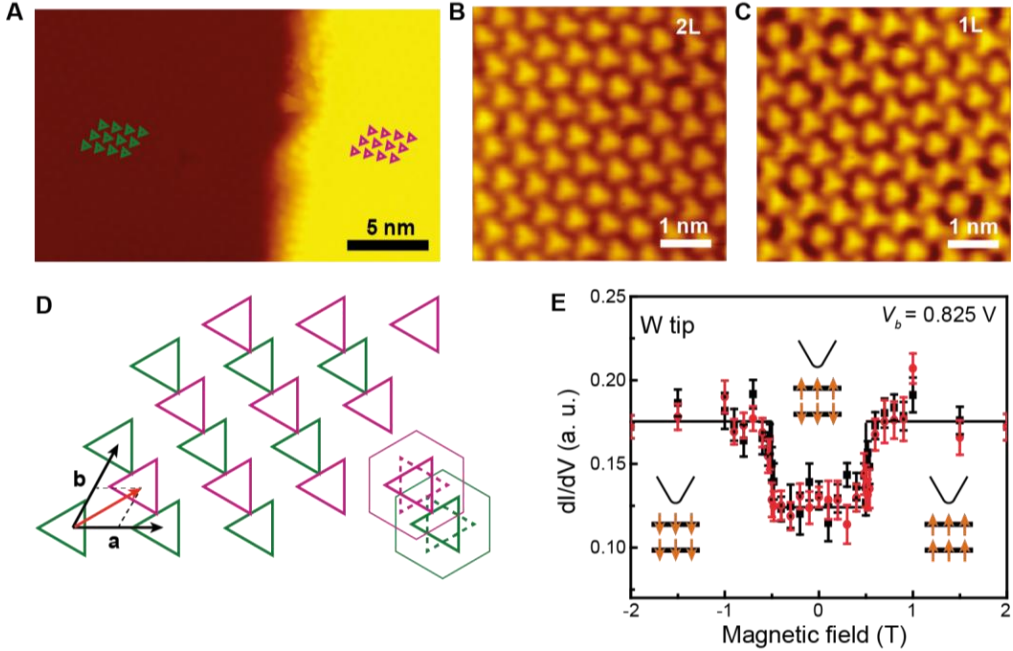

**Fig. 4. Interlayer antiferromagnetic coupling in an R-type stacked CrBr₃ bilayer. A,** STM image of a CrBr₃ film with both a monolayer (1L) region and a bilayer (2L) island. **B, C,** Atomically resolved images of bilayer (**B**) and its extended bottom monolayer (**C**). $V_b$ = 1.9 V. The stacking configuration in the bilayer is identified as R-type, i.e. the top and bottom layers are aligned to the same orientation. **D,** Atomic structure of the CrBr₃ bilayer, as determined from atomically resolved STM scans (see Supplementary Materials Fig. S3 for further details). The representation style follows that in Fig. 3. The unit cell of the top layer (magneta) is translated to 0.48**a** + 0.48**b** of the bottom layer (green). **E,** Spin-dependent tunneling on the CrBr₃ bilayer with a non-magnetic W tip at $V_b$ = 0.825 V. Abrupt increase of the dI/dV signal was observed at magnetic fields of about ±0.5 T, suggesting an interlayer antiferromagnetic coupling within ±0.5 T. The insets show the magnetization arrangements of the bilayer at different magnetic fields.